\documentclass[preprint]{elsarticle}

\usepackage[ ]{algorithm2e}
\usepackage{subcaption}
\usepackage{color}

\usepackage{amsmath,amssymb}

\usepackage{lineno,hyperref}

\begin{document}

\begin{frontmatter}

\title{Cellular non-nonlinear network model of microbial fuel cell}


%
%

\author[label1]{Michail-Antisthenis Tsompanas\corref{cor1}}
\address[label1]{Unconventional Computing Centre, University of the West of England, Bristol BS16 1QY, UK\fnref{label1}}
\cortext[cor1]{Corresponding author}
\ead{antisthenis.tsompanas@uwe.ac.uk}

\author[label1]{Andrew Adamatzky}

\author[label2]{Ioannis Ieropoulos}
\address[label2]{Bristol BioEnergy Centre, University of the West of England, Bristol BS16 1QY, UK\fnref{label2}}

\author[label1]{Neil Phillips}

\author[label4]{Georgios Ch. Sirakoulis}
\address[label4]{Department of Electrical and Computer Engineering, Democritus University of Thrace, Xanthi 67100, Greece\fnref{label4}}

\author[label2]{John Greenman}


\begin{abstract}
A cellular non-linear network (CNN) is a uniform regular array of locally connected continuous-state machines, or nodes, which update their states simultaneously in discrete time. A microbial fuel cell (MFC) is an electro-chemical reactor using the metabolism of bacteria to drive an electrical current. In a CNN model of the MFC, each node takes a vector of states which represent geometrical characteristics of the cell, like the electrodes or impermeable borders,  and quantify measurable properties like bacterial population, charges produced and hydrogen ions concentrations. The model allows the study of integral reaction of the MFC, including temporal outputs, to spatial disturbances of the bacterial population and supply of nutrients. The model can also be used to evaluate inhomogeneous configurations of bacterial populations attached on the electrode biofilms.
\end{abstract}


\end{frontmatter}


%
%
%
%
%

\section{Introduction}

Microbial Fuel Cells are renewable bioelectrochemical transducers that convert biochemical energy into electricity. MFCs empower simultaneous treatment of waste-water and energy extraction from mixed organic media via the usage of microbial consortia as bio-catalysts. In addition to treating anthropogenic waste and wastewater, whilst producing rather than consuming electrical energy, MFCs have the ability of degrading toxic pollutants and the advantage of not burdening further the carbon cycle in the way fossil fuels do \cite{ieropoulos2013miniature}. They bear some resemblance to conventional fuel cells, given that they comprised two compartments, the anode and the cathode, divided by a Proton Exchange Membrane (PEM), where oxidation and reduction reactions occur. A major difference is that MFCs use abundant, renewable fuels, such as organic substrates, that are metabolised by bacteria, whereas chemical fuel cells are fuelled by pure compounds (which could be toxic as methanol or explosive as hydrogen) oxidised by precious metals. As a result, their inexpensive functionality and maintenance designate MFCs as a viable solution for producing energy in isolated areas.

Despite the fact that MFCs were proposed more than a century ago \cite{Potter1911}, it is still a subject of rigorous research, due to the increased power densities that have been achieved in the last decade. In addition, since they operate in ambient conditions (ambient environment temperature, atmospheric pressure, neutral pH) and given the aforementioned advantages, they can efficiently support systems for applications like remotely deployed sensors and robotics \cite{Ieropoulos2016}. Nonetheless, their analysis in copious conducted experiments reveal significant limitations on their performance due to low microbial activity (low growth rate or metabolic rate, due to non-optimal growth conditions or unsuitable microcosm --- insuficient anodophiles), ohmic losses, mass transfer limitations on the electrode surfaces, non-optimised electrode architectures and transfer potential through the PEM. The process of defining the factors that limit the performance of MFCs can lead to more efficient designing methods. Although some techniques used in the conventional chemical fuel cells could be adopted, they can not be expected to provide the same results due to the fundamental differences of the systems particularly due to the biological nature.
 
MFCs are complicated devices that contain bio-electrochemical reactions, mass and charge balance principles, biotic or abiotic transformation processes. As a result, their analysis and design process require a multidisciplinary approach with background in electrochemistry, microbiology, physics and engineering. On top of that, there are numerous differentiations in the MFCs studied which range from their configurations (having two chambers separated by a PEM or a membrane-less single-chamber MFCs) to the type of their incorporated mechanism of donating electrons to the anode electrode (mediated or mediator-less MFCs).

Given the complexity of these systems, the number of parameters that affect their outputs and the costs in time and money needed to perform laboratory experiments, the development of computerised mathematical models simulating these systems is of great importance. The implementation of modelling techniques can contribute to the investigation of the principles covering their operation and affecting their performance, producing better arrangement designs of MFCs and working circumstances.

To address spatial dynamics of biophysical processes in a MFC we designed a cellular non-linear network (CNN) model. A CNN is a uniform regular array of locally connected continuous-state machines, or nodes, which update their states simultaneously in discrete time~\cite{chua1993cnn, chua1998cnn}. Essentially, CNN is a finite-difference scheme with time step 1. A CNN is a subset of cellular automata (CA). A CA is the same as CNN but states of nodes are discrete. CA and CNN are often mixed, many researchers do not differentiate these two types of machines. What is imitated in CA can be imitated in CNN and both offer powerful modelling capability as well. There are several studies published on CNN/CA or hybrid models of reaction-diffusion~\cite{dab1989cellular, berryman1989simulation, chopard1991cellular, chopard1994reaction, chopard1995cellular, droz1997cellular, bandman1999comparative, weimar2002three, matsubara2004reaction, suzuki2005striped} and prototyping of chemical computers~\cite{adamatzky2006phenomenology, adamatzky2006glider, marchese2002directional, adamatzky2005reaction, adamatzky2008evolving, adamatzky2012reaction} and molecular computers~\cite{hiratsuka2001model}, spatial dynamics of bacterial colonies~\cite{picioreanu1996modelling, krawczyk2003nonlinear, vitvitsky2016cellular, odagiri2009threshold, margenstern2013bacteria}. Other approaches relevant to CNN modelling of MFCs are CA models of fuel loading patterns in nuclear reactors~\cite{fadaei2009new}, dynamics of nuclear reactors~\cite{boroushaki2005simulation, hadad2007application, pirouzmand2016simulation, akishina2005cellular}, neutron transport~\cite{hadad2008cellular}, waste water treatment by aerobic granules~\cite{benzhai2014simulation}, sequencing batch reactor~\cite{zhang2006cellular}, fluid flow in a porous medium~\cite{bandman2011using}.

The proposed CNN-based model simulates biochemical and electrochemical reactions in a MFC based on synthetic redox mediators. To the best of the authors' knowledge this is the first attempt to simulate the outputs of a MFC with a CNN model. Despite the fact that the application of CNNs have been widely used to simulate several biological, chemical and physical processes as indicted previously, the novelty of this study can be pinpointed in the fact that all the processes and, thus, the behavior of a batch-fed MFC has not been previously presented in a single CNN lattice with a local state comprised of all the critical quantities (several chemical species concentrations, biomass concentrations and current produced). Such a model will allow for a detailed analysis of integral outcomes of spatial processes inside MFCs, including a possible uneven distribution of nutrients in the MFC chamber, patterns of bacterial population in biofilms covering electrodes and distributions of diffusing metabolites. Nonetheless, the use of CNN as the mathematical basis for the model allows the employment of the inherent fully parallel nature of synchronised locally interconnected simple unities. The subject device is a two chamber MFC with the presence of electroactive microbes in suspension in the bulk liquid and forming biofilms on the surface of a planar anode electrode and assuming electron transfer from the microbes to the electrode with the use of an externally added diffusible chemical mediator. That mechanism can be differentiated to emulate various types of MFCs, that will be the aim of future works. Nonetheless, the present study can lead towards exploiting the parallelism of the simulating tools and, as a result, intensively accelerating the simulation of MFCs, by the implementation of the CNN-based algorithm on parallel hardware, as illustrated in \cite{Dourvas2015,Tsompanas2016}.


\section{Previous work}
\label{sec:sec0}

Despite the intense investigation in the laboratory experimental field to optimise the performance of MFCs, results from computational models are not derived with the same rate. Moreover, the few models developed are targeting specific MFC configurations each and are so strictly specified that they become impractical for implementation on different configurations. The first model presented \cite{Zhang1995}, investigated a single population using an external mediator as receptor of electrons. That model analysed the correlation of the concentration of the external mediator with the higher possible power output.

The authors of \cite{2007} introduced the simulation of a MFC with an added mediator and several populations of suspended and attached biofilm microorganisms. The model was developed on two or three dimensions providing the resultant current produced by homogeneous or not biofilms. The results were derived by taking into account several parameters, like the content of different microbial species, the amount of suspended microbes compared with ones attached, the potential of the mediator, the initial concentrations of the mediator and the substrate and many more. The results provided were compared with experimental data from a batch MFC fed with acetate and inoculated with Geobacter and found in a good agreement.

The model presented in \cite{2007} was updated in \cite{2008} with the incorporation of International Water Association (IWA's) anaerobic digestion model (ADM1) \cite{Batstone2002}. The coexistence of several types of methanogenic and electroactive bacteria is simulated, taking into account whether they are suspended or attached to the anode electrode. A batch MFC was simulated to test the effects of the electrical circuit on the population of the microorganisms and the results were compared with laboratory data. The model is also based on a one, two or three dimensions 
partial differential equations system to represent the spatial distribution of solutes in the biofilm.

In \cite{2010a}, a model simulating a MFC with only suspended microorganisms and externally inserted mediator was studied. In this model the conservation of mass for the dissolved ingredients has a basic role. The biomass growth is not studied in that model while a batch mode MFC which was periodically fed was simulated. The results were compared with laboratory data from MFCs inhabited with suspended \textit{Proteus} cells and incorporating thionine as a mediator, proving accurate representation of the system. The model introduces the ideas of endogenous metabolism or intracellular substrate storage to justify a small amount of current present between the feeding pulses. As the conditions and concentrations in the anode were considered uniform, a one dimensional solution of the algorithm's equations was presented. Some key parameters of the equations used were extracted from fitting the outputs to the experimental data, while some others were estimated.

Another study \cite{2010b} proposing a mathematical model for MFCs, was based on two dimensional macroscale mass balance equations and microscale biofilm evolution. The model contains hydrodynamic calculations and mass and charge balances through diffusion, convection and electromigration to simulate the current output, species concentrations and pH distributions throughout the anode. Nonetheless, the possibility of depicting on two or three dimensions irregular biofilm and electrodes configurations and simulate the effects on the MFC operation and outputs was provided. The model was used to reproduce the system's outputs such as its pH distributions, the effect of multiple communities of electroactive, methanogenic and fermentative bacteria existing in the anode biofilm and the effect of the flow over or through complex electrodes.

The authors of \cite{2010c} presented a model simulating a MFC with inspiration of models simulating chemical fuel cells. A mediator-less two-chamber configuration was studied in steady and dynamic states, with the combination of biochemical reactions, Butler–Volmer expressions, mass and charge balance equations. Also, given the assumption that the anode is under anaerobic conditions the ADM1 was used to simulate the biochemical reactions. The authors argue that the effects imposed by the reactions in the cathode on the performance of the MFC are noteworthy. There were some laboratory experiments conducted in the context of this study to determine some of the parameters required by the model. Despite that, it was not possible to extract some parameters that were estimated by a mathematical method of best fitting the results of the model with the experimental data.

A one dimensional model 
of a steady state MFC incorporating charge, heat and mass transfer and biofilm formation was suggested \cite{2013}. The model simulates the electrical performance of the MFC and the evolution of the biofilm formation that are affected by the inputs, like temperature and substrate concentration. The configurations simulated were the same as in \cite{2010c}, however a minor improvement in the results' agreement was realised mainly because of the addition of the effects that heat transfer has on the anode and the cathode. 

The authors of \cite{2010d} presented a model that concentrates on the effects that the operational parameters of a MFC, such as the external ohmic resistance and organic loads, have on communities of anodophilic and methanogenic microbes. The ordinary differential equations that are included by the model were subjected to fast numerical solution techniques in order for the model to be more efficient than the one presented in \cite{2007} but not oversimplified as the one presented in \cite{Zhang1995}. Moreover, the model was calibrated with the analysis of two single-chamber membrane-less air–cathode MFCs and then validated on two other MFCs.

Another type of MFCs, namely the membrane-less single-chamber MFC, with molasses as fuel was modelled in \cite{2014}. The model consisted of a system of differential equations depicting the diffusion and concentration profiles of the molecules, which are solved with a numerical approximation technique, namely the implicit finite difference method. Despite the fact that the authors find the results of the model in accordance with the theoretical principles, they suggest that more detailed evaluation of the model is required in order to serve the optimisation of the performance of MFCs.

The novel model based on CNN simulating the performance of MFCs proposed here, was motivated by the limitations imposed by the previously reported models. Some of these limiting factors are 
the representation of the MFC compartments in one dimension disregarding possible inhomogeneities in the electrodes or the biofilms and the implementation of complicated differential equations that make models time consuming and compute-intensive. Moreover, the use of CNN is advantageous compared with past models, as complicated computations are emerging from synchronised local interactions of basic simplistic entities. The given homogeneity of CNNs and the local interconnections are essential features for mapping great amounts of cells in digital circuit systems, taking full advantage of a parallel implementation of the model. 

\section{The proposed model}
\label{sec1}

The model proposed here is simulating the performance of a two-chamber acetate-fed MFC with a biofilm attached on the anode electrode in batch mode. The following assumptions are made:

\begin{enumerate}
\item The electrode is considered to be a 2D solid, which although is in contrast with best practices \cite{santoro2015cathode}, it is a necessary simplification for the model. Important characteristics such as low biofilm populations --- due to the limited available outer area for colonisation --- and biofilm erosion will be considered in future studies.
\item The microbial growth is proportional to the spatial concentrations of growth-limiting factors like acetate and oxidised mediator; however the metabolic products formed are not considered as primary factors effecting the performance of the MFC and, thus, neglected by the model;
\item 	An ideal PEM is considered, allowing only protons to pass through but nothing else (like other cations, carbon dioxide, acetate and oxygen) and does not limit the performance of the MFC;
\item 	An added mediator mechanism is used; as a result, electrons are shuttled from the microbes to the anode electrode by the reduced form of a chemical which is oxidised at the surface of the electrode;
\item 	The heat generation by electrochemical reactions occurring on both electrodes, anode compartment and biofilm and the heat flux is not considered;
\item 	Temperature is considered fully controlled and kept constant;
\item 	Solute materials are transported in space through molecular diffusive forces and migration of charged molecules, due to the electrical potential field. The latter is not studied in the current version of the model;
\item 	No gravity forces were included for any of the material to keep the model as straightforward as possible;
\item 	The anode compartment is defined as a continuously stirred tank reactor;
\item 	The reactions in the cathode side are prescribed as constant and are not a limiting factor of the MFC's performance.
\end{enumerate}

Note here, that neglecting the migration of charged molecules in the electrical potential field generated by the voltage outputs of the MFC can not be treated as an oversimplification of the model. That assumption stands, given that the effects of that phenomenon are minor due to the low voltages produced by the system and the high conductivity of the anolyte solution. Moreover, previously published models of MFCs \cite{Ou201649} suggested that the results with and without the incorporation of the phenomenon of electrical migration provides results with an average difference in power densities of 1.92\% throughout the range of MFC voltage outputs. Nonetheless, most of the previous work done on the field of MFC modelling neglected that phenomenon \cite{2007,2010c,2013}.

The grid of the CNN model is $n\times{n}$ (here is consisted of $68\times68$ cells for illustration reasons), and is representing a cross section of an area in the anode compartment near the electrode of a given batch-mode MFC. The size of the simulated area, represented by one CNN cell is set with the actual geometrical distances in mind. The size of each cell has been chosen to be sufficient to illustrate an area where the abstraction of homogeneous reactions occurring can be justified and fluxes of soluble chemicals can be depicted. Namely, the area of a CNN cell is defined as a 1{$\mu{m}$} $\times$ 1{$\mu{m}$} area of the anode compartment, an area comparable with the typical dimensions of some species of bacteria. As a result, the CNN cell is small enough to accommodate the two dimensional projection of the existence of a single bacterium in a the specified area. The size of a CNN cell can be trivially increased after the appropriate decrease of the dimension-related parameters and without changing anything in the algorithmic configurations of the model. This procedure would effect negatively the accuracy of the model, as a higher level of abstraction will be assumed, however the execution time will be decreased as a smaller amount of elements will be calculated.

The von Neumann neighbourhood was used, meaning that each cell has five neighbours, the four closer adjacent cells to the central one (located in the north, south, west and east directions), including the central one. Note here that while the Moore neighbourhood could be used to provide more realistic results (as a more extended neighbourhood would be used including nine cells), the von Neumann neighbourhood was chosen to maintain an efficient point in the trade-off between complexity and accuracy. While the accuracy of the calculations is slightly reduced and the complexity of the model is reduced, its speed of execution is considerably enhanced. Each cell is defined by its state consisted of several parameters that are simultaneously updated throughout the simulation time steps, according to the states of its neighbours and by the local rule. The parameters consisting the state of each cell are the following:

\begin{equation} \label{eq1}
\begin{split}
S^t_{i,j} =[T^t_{i,j}, {X}^t_{i,j}, {(C_{Ac})}^t_{i,j}, {(C_{Mred})}^t_{i,j}, {(C_{Mox})}^t_{i,j}, {(C_{H})}^t_{i,j}, O^t_{i,j}, I^t_{i,j}]
\end{split}
\end{equation}

\noindent where $i$ and $j$ are the dimension indexes that establish the location of each cell in the grid and $t$ is the current time step. $T^{t}_{(i,j)}$ is illustrating the way the relative area, represented by each cell, can be classified depending on its characteristic structure. This parameter can change through the simulation time steps and can have the following values:

\begin{equation} \label{eq2}
T^t_{(i,j)} =
\begin{cases}
		0, \mbox{for borders (confining the movement of elements),}  \\
		1, \mbox{for anode electrode surface,} \\
		2, \mbox{for biofilm attached to the electrode,} \\
		3, \mbox{for bulk liquid of anolyte.} \\
\end{cases}
\end{equation}

\noindent Parameters $X^t_{(i,j)}, {(C_{Ac})}^t_{i,j}, {(C_{Mred})}^t_{i,j}, {(C_{Mox})}^t_{i,j}$ and ${(C_{H})}^t_{i,j}$ indicate the concentrations of ingredients that participate in the significant reactions in a MFC which are investigated for designing the model. Note here, that the units of all the parameters in the following equations and their values for an example configuration presented in the following section can be found in Table \ref{tab:table1}. 
Namely,
\begin{itemize}
  \item $X^t_{(i,j)}$ represents the bacteria populations present in the anode compartment either in suspension or constituting a biofilm on the electrode,
  \item ${(C_{Ac})}^t_{i,j}$ represents the electron donor material for the microbes (the fuel of the MFC), here assigned as acetate, 
  \item ${(C_{Mred})}^t_{i,j}$ represents the added mediator chemical in reduced form, while 
  \item ${(C_{Mox})}^t_{i,j}$ represents the added mediator chemical in oxidised form,  
  \item ${(C_{H})}^t_{i,j}$ represents protons/hydrogen ions that are released from the reactions occurring in the anode electrode surface.
\end{itemize}

Parameters $O^t_{i,j}$ and $I^t_{i,j}$ represent the locally imposed over-potential and produced current density in each cell $(i,j)$. The $O^t_{i,j}$ parameter studied is the activation over-potential, while the effects from the ohmic and concentrations over-potentials are implemented in the local rule calculations \cite{2007}.

The definition of the anode compartment as a continuously stirred tank reactor is based on the fact that phase mixture is quite faster than the electrochemical and biochemical reaction rates \cite{2010c}. As a result, all cells representing an area of anode bulk liquid will obtain the same state at the same time step. Also, the cells representing borders, theoretically undergo no changes throughout time, thus, their states are regulated to a constant set of values, equal to zeros.

The local rule used in the model simulates the kinetics and reactions of all materials. The calculations depend on the type of the cell represented ($T^t_{(i,j)}$) at the given time step and employ the concentrations of chemicals in the predefined area (the neighbourhood of each cell) to provide the concentrations throughout time for each CNN cell. The concentrations within cells simulating the biofilm is given by the following expression:

\begin{equation} \label{eq3}
\begin{split}
{(C_{Ac})}^{t+1}_{i,j} = & {(C_{Ac})}^t_{i,j} + D_{Ac} \times ( {(C_{Ac})}^t_{i-1,j} + {(C_{Ac})}^t_{i+1,j} + {(C_{Ac})}^t_{i,j-1} + {(C_{Ac})}^t_{i,j+1} \\&-  N_{i,j}\times{(C_{Ac})}^t_{i,j} ) - {(rf_{Ac})}^t_{i,j}
\end{split}
\end{equation}

\noindent where $D_{Ac}$ is the diffusion coefficient of acetate, $N_{i,j}$ is the number of available neighbour cells of the cell $(i,j)$ (not borders or the electrode) and ${rf_{Ac}}^t_{i,j}$ represents the rate at which the acetate is consumed by bacteria. This consumption rate depends on the local concentrations of acetate, biomass and oxidised mediator and is defined as a double Monod limitation kinetic equation \cite{OrtizMartinez201550}:

\begin{equation} \label{eq4}
{(rf_{Ac})}^t_{i,j} =  Q_{Ac}  \times {X}^t_{i,j} \times \frac{({C_{Ac})}^t_{i,j}} {K_{Ac} + {(C_{Ac})}^t_{i,j}} \times \frac{({C_{Mox})}^t_{i,j}} {K_{Mox} + {(C_{Mox})}^t_{i,j}}
\end{equation}

\noindent where $K_{Ac}$ is the Monod half-saturation coefficient for acetate, $K_{Mox}$ is the Monod half-saturation coefficient for the oxidised mediator and $Q_{Ac}$ is the maximum specific rate constant for microbial consumption of acetate.

Equation \ref{eq3} is actually the numerical approximation of the implicit finite difference method (eq. \ref{eq3a}) of Fick's second law in two dimensions \cite{2014}, with the addition of the consumption rate of acetate by bacteria. Fick's second law (eq. \ref{eq3b}) is used to simulate the transport of solutions or liquids in other liquids, states that the change through time of concentration is depending on the differences of concentration and is derived from mass balance principle of a species in a fluid continuum \cite{Balluffi2005}.

\begin{equation} \label{eq3a}
{\frac{{{d}^{2}f}}{{dx}^{2}}} \bigg|_{x}= \frac{f(x+\Delta x) -2f(x) +f(x-\Delta x)} {\Delta x^{2}}
\end{equation}

\begin{equation} \label{eq3b}
\frac{\partial c}{\partial t} = D\frac{{\partial}^{2} c}{{\partial x}^{2}}
\end{equation}

\noindent where $c$ is the concentration of a species, $D$ is the diffusion coefficient, $t$ is time and $x$ is the length.

Furthermore, the concentration of acetate in the bulk liquid is equal for all cells, as mentioned previously, due to the assumption of the anode as a continuously stirred tank reactor. Also, biofilm development, along with substrate consumption rate, and solute mass transport, through diffusion, occur at different time scales, in the order of hours and seconds, respectively. Consequently, an immediate effect of the substrate consumption in the biofilm on the bulk concentration will be demonstrated in the mathematical formulas. Taking all the above into consideration, the concentration of acetate in the bulk liquid is given by the following equation:

\begin{equation} \label{eq5}
\begin{split}
{(C_{Ac})}^{t+1}_{i,j} =  {(C_{Ac})}^t_{i,j} + {(rb_{Ac})}^t_{i,j} + \frac{\sum{({rf_{Ac})}^t_{k,l}}} {V_a^t} + \frac{\sum{({re_{Ac})}^t_{m,n}}} {V_a^t}
\end{split}
\end{equation}

\noindent where ${(rb_{Ac})}^t_{i,j}$ is the net rate of reactions in the bulk, $\sum{({rf_{Ac})}^t_{k,l}}$ is the overall reaction rates in the whole biofilm ($ \forall k,l$ that $T^t_{(k,l)} =2$), $\sum{({re_{Ac})}^t_{m,n}}$ is the overall reaction rates in the surface of the electrode ($ \forall m,n$ that $T^t_{(m,n)} =1$) and $V_a$ is the volume of the bulk liquid, here measured by cells that have an area of $1{\mu}m^2$.

The rates of reactions in the bulk liquid (${rb_{Ac}}^t_{i,j}$) are calculated the same way as the rates of reactions in the biofilm (${rf_{Ac}}^t_{k,l}$), specifically by eq. \ref{eq4} with the use of local concentrations for the bulk liquid.

Similar rules apply for the rest of chemical species that are important in the functionality of a MFC, namely the mediator in reduced and oxidised form and hydrogen ions. The following equations describe the evolution of the concentrations in the biofilm region:

\begin{equation} \label{eq5a}
\begin{split}
{(C_{Mred})}^{t+1}_{i,j} = & {(C_{Mred})}^t_{i,j} + D_{M} \times ( {(C_{Mred})}^t_{i-1,j} + {(C_{Mred})}^t_{i+1,j} + {(C_{Mred})}^t_{i,j-1}+ 
\\&  {(C_{Mred})}^t_{i,j+1} -  N_{i,j}\times{(C_{Mred})}^t_{i,j} ) + Y_{M} \times {(rf_{Ac})}^t_{i,j}
\end{split}
\end{equation}

\begin{equation} \label{eq5b}
\begin{split}
{(C_{Mox})}^{t+1}_{i,j} = & {(C_{Mox})}^t_{i,j} + D_{M} \times ( {(C_{Mox})}^t_{i-1,j} + {(C_{Mox})}^t_{i+1,j} + {(C_{Mox})}^t_{i,j-1}+  \\& {(C_{Mox})}^t_{i,j+1} - N_{i,j}\times{(C_{Mox})}^t_{i,j} ) - Y_{M} \times {(rf_{Ac})}^t_{i,j}
\end{split}
\end{equation}

\begin{equation} \label{eq5c}
\begin{split}
{(C_{H})}^{t+1}_{i,j} = & {(C_{H})}^t_{i,j} + D_{H} \times ( {(C_{H})}^t_{i-1,j} + {(C_{H})}^t_{i+1,j} + {(C_{H})}^t_{i,j-1} + {(C_{H})}^t_{i,j+1} - \\& N_{i,j}\times{(C_{H})}^t_{i,j} ) + Y_{H} \times {(rf_{Ac})}^t_{i,j}
\end{split}
\end{equation}

\noindent where $D_{M}$ and $D_{H}$ are the diffusion coefficients of the mediator in reduced and oxidised form and the hydrogen ions, respectively. $Y_{M}$ and $Y_{H}$ are the yield of mediator in both forms and protons from the acetate substrate, respectively.

The concentration of the aforementioned chemical species in the bulk liquid are calculated by the following:

\begin{equation} \label{eq5d}
\begin{split}
{(C_{Mred})}^{t+1}_{i,j} =  {(C_{Mred})}^t_{i,j} + Y_{M} \times {(rb_{Ac})}^t_{i,j} + \frac{\sum{[ Y_{M} \times ({rf_{Ac})}^t_{k,l} ]}} {V_a^t} + \frac{\sum{({re_{Mred})}^t_{m,n}}} {V_a^t}
\end{split}
\end{equation}

\begin{equation} \label{eq5e}
\begin{split}
{(C_{Mox})}^{t+1}_{i,j} =  {(C_{Mox})}^t_{i,j} - Y_{M} \times {(rb_{Ac})}^t_{i,j} + \frac{\sum{[ - Y_{M} \times ({rf_{Ac})}^t_{k,l} ]}} {V_a^t} + \frac{\sum{({re_{Mox})}^t_{m,n}}} {V_a^t}
\end{split}
\end{equation}

\begin{equation} \label{eq5f}
\begin{split}
{(C_{H})}^{t+1}_{i,j} =  {(C_{H})}^t_{i,j} + Y_{H} \times {(rb_{Ac})}^t_{i,j} + \frac{\sum{[ Y_{H} \times ({rf_{Ac})}^t_{k,l} ]}} {V_a^t} + \frac{\sum{({re_{H})}^t_{m,n}}} {V_a^t}
\end{split}
\end{equation}

It must be mentioned here that the reaction rates on the electrode surface for the model are designed having in mind the oxidation mechanism of the mediator and the fact that the electrode stands theoretically as an impermeable border for all the other solute materials. As a result the reaction rate for the acetate substrate is equal to zero ($({re_{Ac})}^t_{i,j}=0$), whilst, for the both mediator forms is depending on the produced current density and calculated as:

\begin{equation} \label{eq5g}
\begin{split}
({re_{Mred})}^t_{i,j}= - \frac{I^t_{i,j}}{2F}
\end{split}
\end{equation}

\begin{equation} \label{eq5h}
\begin{split}
({re_{Mox})}^t_{i,j}= \frac{I^t_{i,j}}{2F}
\end{split}
\end{equation}

The reduction reactions that involve the hydrogen ions, are occurring in the cathode electrode, which is not studied in detail in the present study. Thus, in order to simulate the decrease in the concentration of hydrogen ions in the anode compartment, which is due to the flux of the ions towards the cathode electrode, an upper limit was imposed in all the anode area. If the concentration of hydrogen ions exceeds a predefined value (namely $C_{H_{max}}$) anywhere in the anode, its value will be fixed to the limit.

The over-potential imposed locally on the surface of the electrode is based on the concentrations of protons, reduced and oxidised mediator and the current produced.

\begin{equation} \label{eq6}
\begin{split}
{O}^{t}_{i,j} =  {E}_{c} - R_{tot} I^t_{i,j} - (E^0_M + \frac{RT}{2F} \ln{\frac{{(C_{Mox})}^t_{i,j} ({(C_{H})}^t_{i,j})^2}{{(C_{Mred})}^t_{i,j}} } ) 
\end{split}
\end{equation}

\noindent where ${E}_{c}$ is the constant value assumed for the cathode potential, $R_{tot}$ is the total resistance (internal and externally connected), $E^0_M$ is the standard reduction potential for the mediator, $R$ is the gas constant, $T$ is the temperature and $F$ is the Faraday constant. 

For the calculation of the current density produced on the anode electrode's surface, the widely used Butler-Volmer equation \cite{2007,newman2004electrochemical} is implemented in the model:

\begin{equation} \label{eq7}
\begin{split}
{I}^{t}_{i,j} = & {I}_{ref} \bigg(\frac{{(C_{Mred})}^t_{i,j}}{{(C_{Mred})}_{ref}}\bigg)\bigg(\frac{{(C_{Mox})}^t_{i,j}}{{(C_{Mox})}_{ref}}\bigg)^{-1}\bigg(\frac{{(C_{H})}^t_{i,j}}{{(C_{H})}_{ref}}\bigg)^{-2}\cdot \bigg(exp{({2.303\cdot{O}^{t}_{i,j}}/{b})}-
\\& exp{(-{2.303\cdot{O}^{t}_{i,j}}/{b})}\bigg)
\end{split}
\end{equation}

\noindent where ${I}_{ref}$ is the exchange current density for mediator oxidation in reference conditions and $b$ is the Tafel coefficient for mediator oxidation (or referred to as the anodic/cathodic Tafel slope, representing the over-potential increase required for a $\times$10 increase in current: $b=\frac{2.303RT}{{\alpha}F}$, where $\alpha$ is the anodic transfer coefficient). 

It is worth-mentioning that the over-potential and current density are meaningful only on the cells representing the electrode's surface. Thus, the aforementioned equations provide these parameters for cells that have $T^t_{(i,j)} =1$, otherwise they are equal to zero. 

As the calculations of the over-potential and the current density are implicit, a random positive value of the total current is assumed and then a numerical method is used to approximate the lowest error in the calculation of the total current, given the concentrations of chemical species on the electrode surface. On the first step, the randomly chosen current value is used (in eq. \ref{eq6}) to calculate the over-potential, which is then used to calculate the new current density (in eq. \ref{eq7}) and, thus, the total current produced. The current density error, which is targeted to be minimised, is given by:

\begin{equation} \label{eq7a}
\begin{split}
e^{t}_{I} = | I^{t}_{new} - \sum{({I}^{t}_{i,j})} |
\end{split}
\end{equation}

\noindent where $I_{new}$ is the randomly assumed value for the first step or the calculated value by the previous step.

The evolution of the biofilm is proportional to the concentrations of chemical species as described by a double Monod limitation kinetic equation (as in eq. \ref{eq4}). The calculation of its expansion in the area of the bulk liquid of the anode is based on a simple algorithm to keep the overall computational model as straightforward and fast as possible. The concentration of biomass material in the biofilm CNN cells is calculated by the following equation:

\begin{equation} \label{eq8}
\begin{split}
{X}^{t+1}_{i,j} =  {X}^{t}_{i,j} + {(rf_{X})}^t_{i,j} 
\end{split}
\end{equation}

\noindent where ${(rf_{X})}^t_{i,j}$ is the rate of the biomass production in the biofilm and equals:

\begin{equation} \label{eq9}
\begin{split}
{(rf_{X})}^t_{i,j}  =  {Y}_{X} \times {(rf_{Ac})}^t_{i,j} 
\end{split}
\end{equation}

\noindent where ${Y}_{X}$ is the biomass yield on acetate substrate. The same equations are used for the approximation of the suspended biomass in the bulk liquid, with the usage of the appropriate local concentrations.

\begin{equation} \label{eq9a}
\begin{split}
{X}^{t+1}_{i,j} =  {X}^{t}_{i,j} + {(rb_{X})}^t_{i,j} 
\end{split}
\end{equation}

\noindent where ${(rb_{X})}^t_{i,j}$ is the rate of the biomass production in the bulk liquid and equals:

\begin{equation} \label{eq9b}
\begin{split}
{(rb_{X})}^t_{i,j}  =  {Y}_{X} \times {(rb_{Ac})}^t_{i,j} 
\end{split}
\end{equation}

An upper limit on the concentration of biomass in the bulk liquid ($X'_{max}$) is imposed to control the biomass growth. Moreover, the following method, which is written in pseudo-code, is performed to simulate the release of pressure produced by the creation of new biomass inside the region of the biofilm. This is to ensure a more realistic expansion of the biofilm attached on the anode electrode towards the bulk liquid. The algorithm is performed with initial cells that are part of the biofilm (for cells $i,j$ that have parameter $T^t_{(i,j)} =2$).

\begin{algorithm}[H]
\SetAlgoLined
  \eIf{${X}^{t+1}_{i,j}>=X_{max}$}{
    \If{a random neighbour cell $(k,l)$ with $T^t_{(k,l)}=3$ can be found}
    {assign $T^{t+1}_{(k,l)}=2$ and ${X}^{t+1}_{k,l}={X}^{t}_{k,l}+{X}^{t}_{i,j} \cdot 0.005$ \;
    Run \textbf{Algorithm 1} for $i=k$ and $j=l$ \; 
    (recursive execution until ${X}^{t+1}_{r,p}<=X_{max}$ )\\
    ${X}^{t+1}_{i,j}={X}^{t}_{i,j} \cdot 0.995$\;}   }{
   ${X}^{t+1}_{i,j}$ is calculated as in eq. \ref{eq8} \;
  }
  
 \caption{Conditions in the evolution of the biomass}
\end{algorithm}

Despite the fact that the model is specifically designed to simulate a MFC in batch mode, a system with continuous flow can be studied by changing the boundary conditions. That approach is the subject of a future study.

\section{Experiments}
\label{sec2}

The configuration of parameters used by the model to produce the results presented in this section, are depicted in Table \ref{tab:table1}. Each time step represents a time period of one tenth of day. The time period of a time step was chosen based on the trade-off between accuracy of the results and execution time. As for smaller time periods the results were not significantly different the execution speed was not burdened with a smaller period of time steps.

\begin{table}[!tb]
\begin{center}
 \begin{tabular}{|| c | c || c | c ||} 
 \hline
 Parameter & Value & Parameter & Value \\ 
 \hline\hline
 ${I}_{ref}$ & $2 \times 10^{-4}$ $A/m^2$ & ${X}^0_{i,j}$ (in bulk liquid) & 0.2 $gCOD / m^3$ \\ 
 \hline
  $b$ & 0.12 $V$ & ${X}^0_{i,j}$ (in the biofilm) & 0.8 $gCOD / m^3$ \\ 
 \hline
  $E_{c}$ & 0.68 $V$ & ${(C_{Ac})}^0_{i,j}$ & 100 $gCOD / m^3$ \\ 
 \hline
  $R_{tot}$ & 100 $\Omega$ & ${(C_{Mred})}^0_{i,j}$ & 0.001 $mM$ \\ 
 \hline
 $E^0_{M}$ & 0.477 $V$ & ${(C_{Mox})}^0_{i,j}$ & 1 $mM$ \\
 \hline
  $R$ & 8.31 $J/(mol \times K) $ & ${(C_{H})}^0_{i,j}$ & 0.001 $mM$ \\ 
 \hline
  $T$ & 298 $K$ & $D_{Ac}$ & $6.5 \times 10^{-6}$ $m^2 / day$\\ 
 \hline
  $F$ & 96485 $C/mol$ & $D_{Mred}$ & $2 \times 10^{-6}$ $m^2 / day$\\ 
 \hline
 $e_{I}$ & 0.00002 $A$
 & $D_{Mox}$ & $1.7 \times 10^{-6}$ $m^2 / day$\\
 \hline
 $K_{Ac}$ & 100 $gCOD / m^3$ & $ D_H $ & $1.16 \times 10^{-6}$ $m^2 / day$\\
  \hline
 $K_{Mox}$ & 0.1  $mM$ & $C_{H_{max}}$ & 0.045  $mM$ \\
 \hline 
  $X'_{max}$ & 17  $gCOD / m^3$ & $X_{max}$ & 18$gCOD / m^3$ \\
 \hline
 $Y_{M}$ & 0.0473 $\frac{(mol mediator)} {(gCOD
acetate)}$ & $Y_{H}$ & 0.0098 $\frac{(mol H+)} {(gCOD
acetate)}$ \\  [1ex]
 \hline
 $Y_{X}$ & 0.243 $\frac{(gCOD biomass)} {(gCOD
acetate)}$ & $Q_{Ac}$ & 10 $\frac{(gCOD acetate)}{(gCOD biomass) \times day} $ \\ [1ex] 
 \hline
\end{tabular}
\end{center}
\caption{Parameters used in the execution of the model.}
\label{tab:table1}
\end{table}

\begin{figure}[!tbp]
    
    \centering
    
        \begin{subfigure}{0.6\textwidth}
        \includegraphics[width=\linewidth]{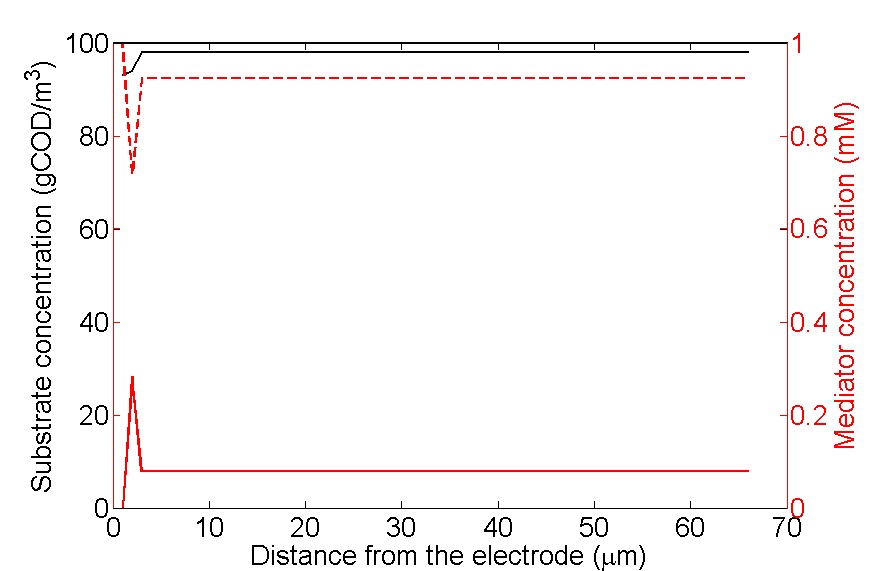}
        \caption{Day 1}
        \label{fig:init1}
        \end{subfigure}

        \begin{subfigure}{0.6\textwidth}
        \includegraphics[width=\linewidth]{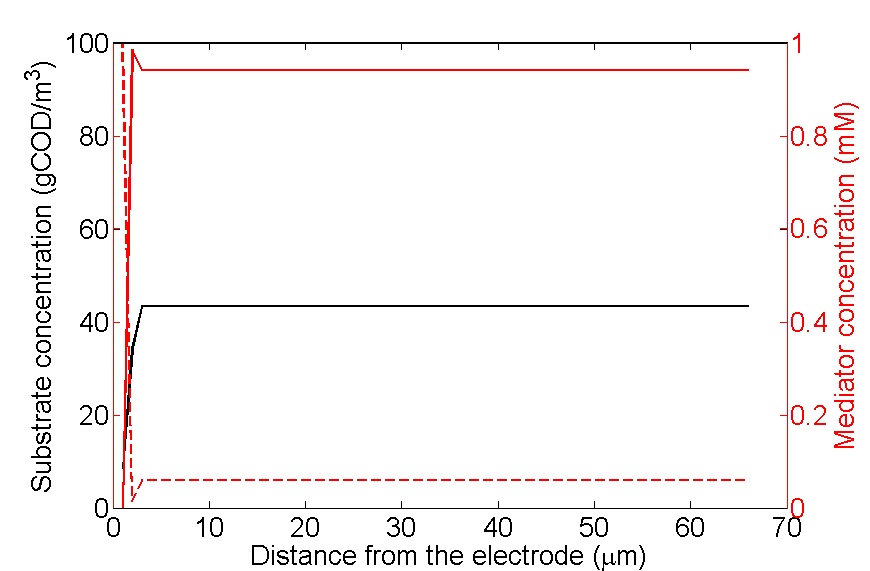}
        \caption{Day 5}
        \label{fig:init2}
        \end{subfigure}
        
        \begin{subfigure}{0.6\textwidth}
        \includegraphics[width=\linewidth]{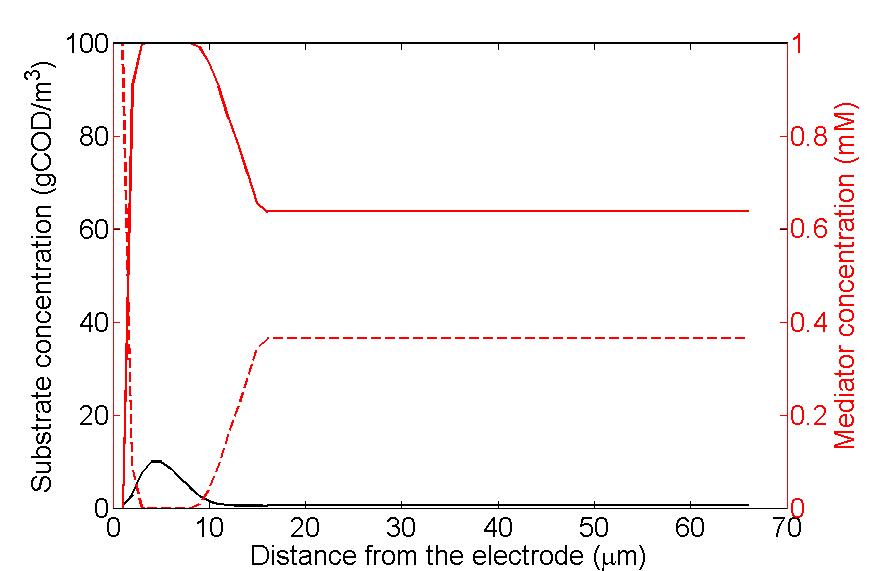}
        \caption{Day 10}
        \label{fig:init3}
        \end{subfigure}

    \caption{The concentrations -- averaged over the columns of cells along the length of the electrode -- of materials along the distance away from the anode electrode during different days (dashed line represents the oxidized form of the mediator). }
    \label{fig:days}
\end{figure}

\begin{figure}[!tbp]
    
    \centering
    
        \begin{subfigure}{0.6\textwidth}
        \includegraphics[width=\linewidth]{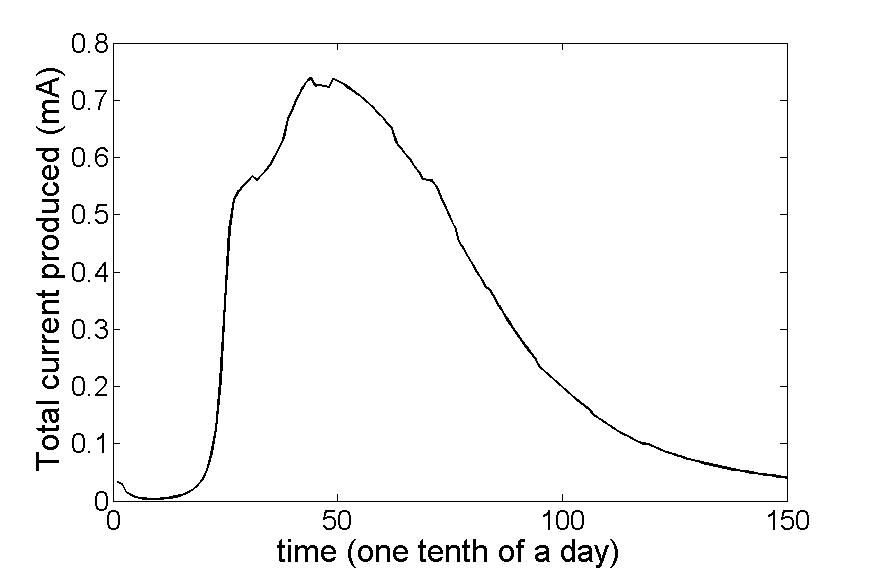}
        \caption{}
        \label{fig:init4}
        \end{subfigure}

        \begin{subfigure}{0.6\textwidth}
        \includegraphics[width=\linewidth]{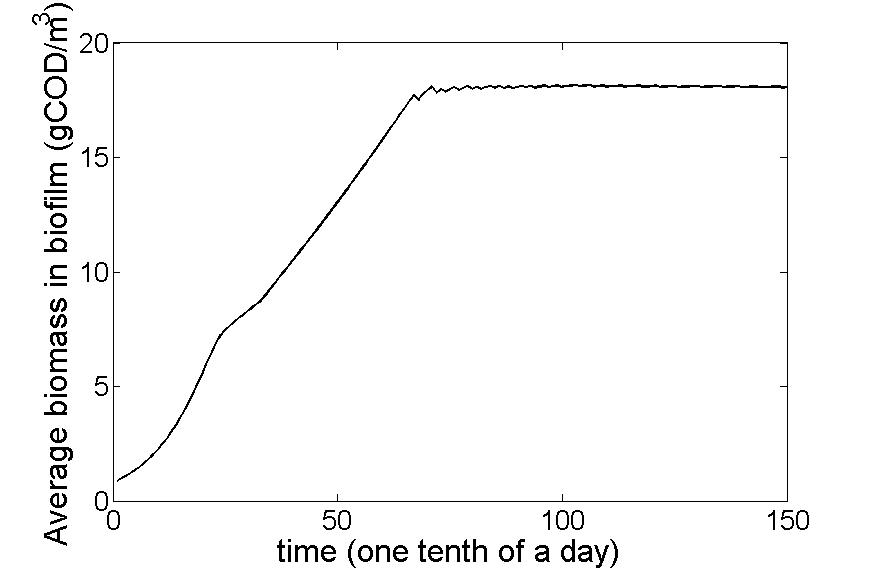}
        \caption{}
        \label{fig:init5}
        \end{subfigure}
        
        \begin{subfigure}{0.6\textwidth}
        \includegraphics[width=\linewidth]{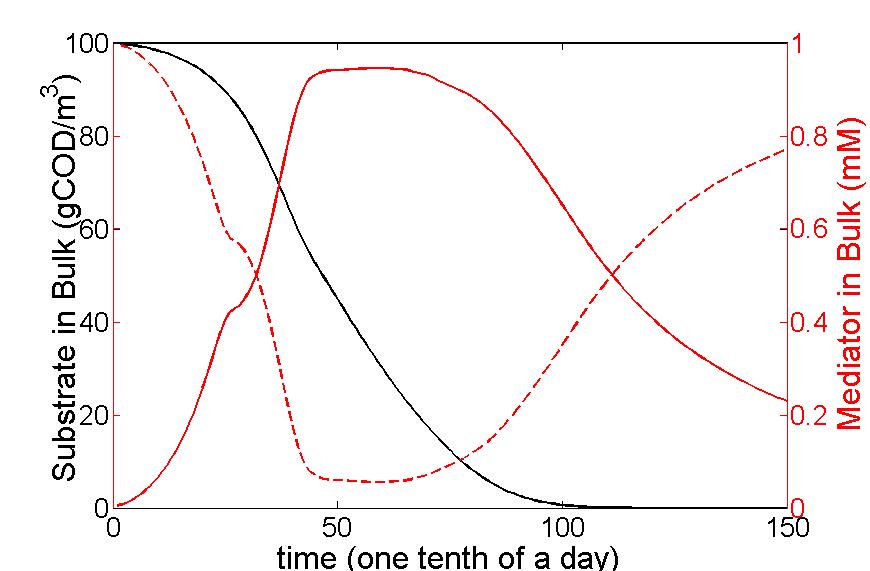}
        \caption{}
        \label{fig:init6}
        \end{subfigure}

    \caption{Model's results: (a) the current output, (b) the evolution of biomass and (c) the evolution of concentrations in the bulk liquid.}
    \label{fig:tot}
\end{figure}

The results derived by the model are illustrated in Figs. \ref{fig:days} and \ref{fig:tot}. Figure \ref{fig:days} depicts how the average --- of the columns of cells along the length of the electrode --- concentrations of materials are changing in one dimension --- illustrated in the $x$-axis of the graphs ---, namely from the electrode moving outwards to the bulk liquid, throughout the time steps of the model that represent the days of the real experiment. Note here that the thickness of the biofilm can be identified as the distance in $x$ axis from the electrode surface position to the point where the stable values of concentrations start. These stable values depict the existence of the continuously stirred tank reactor, namely the bulk liquid. Moreover, Fig. \ref{fig:tot} describes the evolution of the concentrations of every material in the bulk liquid, the current produced and the expansion of the biofilm.

The results depicted in Fig. \ref{fig:tot} are in good agreement with the theoretical principles covering the functionality of MFCs in batch mode with added mediators which are also depicted in previously proposed models investigating similar systems \cite{2007}. It must mentioned here that, despite the biomass initial concentrations, all the other parameters used in previous models \cite{2007} to provide results are the same inserted in the model presented here and illustrated in Table \ref{tab:table1}. Also, the biomass simulation and the limitation of hydrogen ions are simplified compared to the previous models.

The following statements which are easily observed in the results of the model presented in Fig. \ref{fig:tot} are following the general theoretical performance of batch mode MFCs with added mediators. The current output reaches its maximum value within three to four days that retains for a short period (less than a day) and then its value is asymptotically reduced. The concentration of acetate is reduced to zero in the first ten days, while the biomass concentration in the biofilm increases and reaches a plateau when the acetate concentration is low enough. The mediator is a constant quantity that can be found in either oxidised or reduced form, thus adding the concentrations in any time step equals to $1 mM$ which is the initially defined concentration. The oxidised mediator follows the reduction of the acetate substrate, until the current produced is high enough to cause the oxidation reaction in the electrode to overcome the reduction reactions occurring in the biomass. Then, as there is no more acetate to fed reduction reactions, the concentration of oxidised mediator is asymptotically reaching its initial concentration as a product of the oxidation reaction in the electrode. That is also the reason for the asymptotic reduction of the current, namely the lower ratio between reduced and oxidised mediator. Similar explanation of the progress of the oxidised mediator stands for the reduced mediator. 


\section{Discussion} 
\label{sec3}

The advantages of CNN towards other mathematical methods are their simplicity that does not have an effect on their robustness and capability to simulate complicated phenomena, their inherent parallel nature that make them ideal for implementation in contemporary parallel computing devices and their local activity character that combined with the aforementioned features empower their execution on specialised hardware. As a result, the novelty of the model proposed here can be found in the method of numerical approximation, namely CNN, of the equations giving the kinetics of reactions occurring in a MFC and its performance. 
 As mentioned before, CNN is a simple method as it uses a group of simple cells located on a grid, characterised by a state (a set of parameters) and updated based on the same local rule. It has been proved \cite{chua1993cnn} that complex computations can emerge from local interactions of basic entities. Moreover, the synchronised functionality of the basic entities enables a fully parallel execution of computations throughout the grid, accelerating significantly the production of results by the model. In addition to the ability of CNNs to be efficiently executed in parallel computers, their homogeneity, simplicity, synchronised activity and the local characteristic of the interconnections allows an effortless implementation in hardware, aiming further acceleration of the computations \cite{Dourvas2015,Tsompanas2016}. These computing circuits can be cost efficient and pre-manufactured Field Programmable Gate Arrays (FPGAs) and Graphics Processing Units (GPUs) or fully custom, providing higher performance efficiency Application Specific Integrated Circuits (ASICs).  

The present study is based on a time explicit scheme which, despite the fact that might not be as accurate as implicit scheme in general, and especially with large simulation time steps, allows a less complicated implementation and requires a lower computational effort. 
A major advantage of the CNN-based model is that the possible inhomogeneities in the progress of the biofilm or in the structure of the anode electrode can be easily illustrated by the local rule or the initialisation of the cells' states. 

Some models presented previously, simulate the MFC in just one dimension \cite{2013}, oversimplifying the reactions occurring. On the other hand, in \cite{2007,2008,2010b,2014} two dimensional and three dimensional representations are provided that can account for complicated electrode sizes and biofilm formation; however, inhomogeneities would be difficult to implement. The model presented here can easily be scaled to three dimensions and as CNN are the basis of the model, inhomogeneities in the whole volume of the MFC (PEM, electrodes and biofilms) can easily be studied and efficiently recreate a wide variety of actual systems. Despite the fact that an homogeneous area (in the x dimension) was studied here as a proof of concept, the ability of the model to simulate inhomogeneities in two dimensional paradigms is trivial. 



Moreover, in \cite{2007} the competition of two different species, fed on the same substrate used, was investigated, while in \cite{2008} several communities of methanogenic and electroactive bacteria were simulated. The CNN based model is simulating the operation of a MFC inoculated with a single species of bacteria. However, the local dynamics of CNN allow the investigation of more complicated bacterial communities. That is an aspect of an ongoing study.

Nonetheless, some models are designed to calculate significant parameters of a MFC after it has reached its steady state \cite{2013}. Here, the evolution of the outputs of the MFC through time given its inputs is investigated because the process of reaching the steady state is a demanding procedure and should be optimised.



\section*{Conclusions} 
\label{sec4}

The notion of modelling key procedures of complicated processes that occur in real life, enables the scientific community to recreate and study time consuming and expensive laboratory experiments. As a result, scientists can investigate the conditions and parameters of a phenomenon that are difficult to measure in real life. The model proposed here can serve as a virtual lab, which scientists and engineers can utilise to test and justify their theoretical approach to the functionality of MFCs. Moreover, a more efficient designing of these systems can be based on the successful modelling of the processes occurring in MFCs. 

The CNN-based model simulating the performance of a two-chamber acetate-fed MFC described here, is designed in two dimensions and studies cross-section of an area near the anode electrode. The concentrations of chemicals that are involved in the process of producing current and biomass, throughout time, are studied. The local rule of the CNN structure is designed to reflect the double Monod limitation equation, Fick's second law of diffusion and the Butler-Volmer equation.

During the process of designing and evaluating the model the following conclusion was reached. The biofilm's distribution, its initial concentration and the way it evolves and expands through time are greatly affecting the MFC's outputs. Biofilms are difficult to predict and it is challenging to formulate algorithms that simulate their behaviour. That is a major aspect that limits the production of a plethora of MFC models, compared with the ones describing conventional fuel cells. 

The results provided by the model are in good agreement with the theory and the general concepts described in the literature for the functionality of MFCs as well the results of previously published works on MFC modelling. Note that the results produced by the proposed model simulating the performance of the MFC in 15 days is executed in less than a minute in a contemporary computer. Consequently, shifting from two dimensions to three to get more accurate results, given the possible inhomogeneities in MFC compartments or electrodes, will not be prohibited in terms of execution times.

Nonetheless, the fact that the model is based on CNN can be proved further advantageous, as its simplicity, repeatability and local interactions enable its implementation on hardware. That will in turn hugely accelerate the execution of the calculations of the described equations.

The incorporation of the reduction reactions that occur in the cathode side of the MFC and the study of their effects on the performance of the system are aspects of future work. Moreover, multi-species biofilms will be investigated including non- and electroactive bacteria to illustrate the possible competition over the common substrate. Furthermore, the simulation of a MFC under constant flow will be presented with minor alterations in the local rules of the bulk liquid which will be the first step to study systems that are interconnected.

%
%

\section*{Acknowledgements}

This work was funded by the European Union's Horizon 2020 Research and Innovation
Programme under Grant Agreement No. 686585. \\
https://ec.europa.eu/programmes/horizon2020/

\end{document}